\documentclass{article}

\begin{document}

\section{Mass is variable}

	 It is assumed that, for particles having rest mass, mass is equivalent to kinetic energy with coupling constant $c^2$.  This paper  examines the consequences for Newtonian gravity, in flat spacetime, when mass changes according to speed and gravity.  It is assumed that gravitational mass is the same as inertial mass.  The inertial mass increase with speed is:
\begin{equation}
	m = \gamma m_0 = m_0/\sqrt{1-\beta^2}                                           
\end{equation}
where $m_0$ is the rest mass and $\beta=v/c$.  Some physicists dislike this interpretation, preferring to regard rest mass as a constant and refusing to interpret "m" as mass.  It is a matter of taste.  Upon accepting (1), it is clear that different observers will report different masses for a given object, according to their relative velocities. 
	Gravity  leads to changes in $\gamma$, throughout an orbital motion.  In the conventional description of a conservative gravitational system, kinetic energy T is gained at the expense of gravitational potential energy U and the sum of these energies is constant. Thus, when the reference point at which $U=0$ is at $r=\infty$, gravitational attraction leads to a hyperbolic orbit. 
\begin{equation}
	1/\alpha \equiv 1+GM/rc^2 
\end{equation} 
\begin{equation}
	T=(\gamma -1)m_0c^2
\end{equation}
\begin{equation}
	U=(1-1/\alpha)m_0c^2     
\end{equation}
and so, $\gamma=1/\alpha$.  In a closed orbit, where the kinetic energy is insufficient to permit escape, the potential reference point is  some value of r less than infinity, R. The SR gamma boost is modified in that case 
\begin{equation}
	\gamma=1/\alpha_R                                                                      
\end{equation}
\begin{equation}
	1/\alpha_R \equiv 1+GM/rc^2-GM/Rc^2                                                                          
\end{equation}

	In addition to the effect of gravity on speed, which affects the mass, it is postulated that gravitational potential directly affects rest mass.   The Pound-Rebka (ref 1) Mossbauer experiment shows that a falling photon, in a gravitational field, "g", appears to gain energy:
\begin{equation}
	h\nu^\star=h\nu(1+gh/c^2)                                                                               
\end{equation}
The source is at the upper level and the absorber is at the lower level, separated vertically by "h".  It was found necessary to decrease the relative energy of the source by $gh/c^2$ in order to achieve resonance at the absorber. 

	Now, consider a gedanken annihilation experiment.  A thermalized (low kinetic energy) positron-electron pair of total rest mass $m_0$ annihilates at the upper level, and produces two or three photons of total energy $h\nu$. These are reflected, if necessary, downward so they can be compared with the results of a similar annihilation taking place at the lower level.  By energy balance,  photons of somewhat larger energy are produced at the lower level, per (7), if their energy is to equal that produced at the upper level plus the increase due to gravity.  This would mean that the rest mass at the lower level, $m_0^\star$, the source of this larger  energy, is greater than $m_0$ by $m_0gh/c^2$ . 
\begin{equation}
	m_0^\star= m_0(1+gh/c^2)                                                                              
\end{equation}
More generally, "g" changes and one finds that the rest mass increases under gravity:
\begin{equation}
	m_0^\star= m_0/\alpha                                                                             
\end{equation}

The final expression for the mass, in general, is:
\begin{equation}
	m^{\star}=\gamma m_0/\alpha=m_0/\alpha_R\alpha
\end{equation}                                                                                         
Both factors affecting mass, i.e. speed and gravity, are included in (10).  The factor $1/\alpha$ accounts for the increase of rest mass, and the factor $\gamma=1/\alpha_R$ accounts for the SR boost.   The consequences of using this expression for mass in Newton's law of universal gravitation are explored in this paper.

	However, it will subsequently be shown that there are problems with this derivation of (9).  This derivation accepted that photons really change energy, as they fall in a gravitational field.  It will be shown in section 9 that photons move in a gravitational field without change of energy, meaning that any change of energy takes place at the point of origin rather than throughout the path.  However, (9) remains valid as shown by the successes it achieves in dealing with the three classical tests of GR.

\section{Gravitational potential energy}

	Before investigating the changing metric implied by (10), the increase of rest mass (9) leads to a "factor of two" problem with conventional gravitational potential energy.  Normally, one writes this potential energy for $R=\infty$ as
\begin{equation}
	U=-GMm/r
\end{equation}                                                                                         
The gravitational force is taken as the negative gradient of U, and energy conservation is assumed.  This yields the kinetic energy, $1/2\hspace{0.05 in}mv^2$.  It disagrees with (10) where, for a given loss of potential energy, one gains twice as much kinetic energy, T.
\begin{equation}
	T=(m-m_0)c^2= m_0c^2(1/\alpha^2-1)\approx 2GMm_0/r                                          
\end{equation}
The kinetic energy T includes both $1/2\hspace{0.05 in}mv^2$ and $c^2$ times the increase of rest mass.  Since energy is conserved, (11) underestimates U by a factor of two and so the loss of potential energy is actually
\begin{equation}
	U=m_0c^2(1-1/\alpha^2)\approx-2Gm_0M/r                                                        
\end{equation}
The gravitational force is no longer the negative gradient of the entire potential energy, but only of that part associated with velocity gain.

	This expansion of the concepts of the kinetic and potential energy is unconventional, but is forced upon us by the recognition of the rest mass increase in a gravitational potential.  Equations (12) and (13) are critical to an understanding of the nature of gravity.  Conventionally, with potential energy written as (11), the location of the potential energy has been vaguely attributed to "the gravitational field".  Now, we can perhaps be more specific.  In a speculative paper proposing an EM basis of gravity, it was suggested that fractional charges (partons) oscillate within elementary charged particles at the Compton frequency, E/h. (ref 2) This oscillation both creates, and interacts with, long-range 1/r electric fields.  The interaction "heats up" the moving partons, increasing the rest mass of the elementary particle.  These partons are thought to move in a 1/r potential, within the elementary particle, and the virial theorem requires that $\langle T\rangle=-1/2\hspace{.04 in}\langle U\rangle$.  So, any increase of $\langle T\rangle$ within the elementary particle results in a lowering of the total energy.  This, then, is proposed as the storage site for the gravitational potential energy of mass.  Aside from any translational kinetic energy gained by a falling body, the net energy of an elementary particle falls according to the strength of the long-range transient electric fields and leads to an attractive force qualitatively resembling Newton's law of universal gravitation.

\section{Effect of mass on the metric
}

In the presence of a gravitational potential, the metrics of length and time are reduced (as measured) because of the increase of mass.  Recall that the length of a meter stick is equal to some large number times the size of an atom.  The Bohr radius is a measure of the size of an atom, and is inversely proportional to the mass of an electron, "m". (ref 3) 
\begin{equation}
	r_n=n^2h^2\epsilon_0/(\pi mZe^2)    	                                                      
\end{equation}
It follows that the length L* of a meter stick in a gravitational potential is  shortened, by the factor $\alpha/\gamma$.
\begin{equation}
	L^\star=(\alpha/\gamma)L                                                                                           
\end{equation}
This shortening is isotropic, and the metric of distance, as measured, is at any point a scalar function of r.  

	There are several possibilities for measuring time.  An atomic clock is both meaningful and portable.  The term (energy) levels of an atom are proportional to the mass of the electron, and so the splitting of these levels increases when the atom is immersed in a gravitational potential.    In a Bohr atom, (ref 3)
\begin{equation}
	E_n=-mZ^2e^4/(8\epsilon_0^2h^2n^2)                                                                       
\end{equation}
A photon arising from a transition between two of these term levels in a gravity environment appears to have a higher energy than  when the photon is produced in a non-gravity environment.  Expressing the energy of the photon in terms of frequency, and coupling this frequency to the hands of a clock, the result is that the interval between ticks of the clock is shortened by gravity.  The clock runs faster.  
\begin{equation}
	\tau^\star=(\alpha/\gamma )\tau                                                                                                                                              
\end{equation}
It should be emphasized that these changes of mass and of measured length and time, referenced to a gravity-free space, are real.  
	What, then, is the nature of space?  Without measurement, we don't  know.  The problem of measurement is that the calibration of our measuring instruments changes because the instruments  themselves have mass. Meter sticks shrink.  It takes more meter sticks to span a space when that space is influenced by gravity.  Dynamical events are timed differently by co-moving clocks than by a clock sited far away from gravity.  As for the curvature of space, consider a meter stick moved slowly and "flat-on" from a place of low gravitational potential to one of high gravitational potential.  Despite our assumption of flat spacetime, the ends of the meter stick trace curved paths, the ends moving towards one another.  The only "straight" lines are radial, relative to the source of gravity, and even then the metric is nonlinear along the line.

	In summary, then, at the site of a dynamical event, mass is increased by $\gamma/\alpha$, from which it follows that measured lengths are contracted by $\alpha/\gamma$ and time intervals are also contracted by $\alpha/\gamma$.  The philosophy by which these changes occur is quite different than what one assumes in relativity theory.  The outside (non-gravity) observer cannot observe directly these changes.  He infers them.  From his space, he can only "see" the consequences of these changes.  For example, he can send a signal through a region of gravitational potential and find an anomalous time delay for that passage.  Or, he finds a star image deflected when the light path grazes the sun.  Or, he sees a precession of the perihelion of a planetary orbit.

\section{Implications of the changing mass metric
}

	Consider velocity.  What does the observer within a gravitational field measure for the velocity of light?  He reports the velocity, as interpreted by an outside observer,  as 
\begin{equation}
c^\star=(\delta L^\star/\delta\tau^\star)
=(\alpha/\gamma)\delta L/{(\alpha/\gamma)\delta\tau}=c                                                                                                                                         
\end{equation}
Since each element has been contracted by $\alpha/\gamma$, he finds v=c.  What does an outside observer conclude from this experiment?  He cannot follow, in detail, the moving light beam, since his instruments change calibration.  The measurements available to him, i.e. an anomalous time delay, tell him that the speed of light has somehow been reduced in the region near the sun or else the distance has increased.  Despite the fact that the locally measured speed of light is everywhere a constant, c, this theory requires that his "outside" observations reflect that the speed of light at every point along the path is less.  That is, he uses a meter stick within the gravity environment for length measurement, but uses his own clock to measure time.  Since there are no restrictions on the beam of light, it is appropriate that we set  $\gamma=1/\alpha$  and we infer that, so far as the outside observer is concerned, 
\begin{equation}
	c^\star=\alpha^2c                                                                                                                                              
\end{equation}

	In fact, all velocities in a gravity environment are reduced, relative to a non-gravity environment.  What about acceleration?  Referenced to an outside observer's clock, the contraction of distance means that acceleration within gravity is reduced by $\alpha/\gamma$.  This preserves for the outside observer Newton's second law,  since the mass increases by $\gamma/\alpha$ and the product of mass times acceleration remains the same.  Referenced to a clock sited within gravity, however, acceleration is increased by $\gamma/\alpha$.
\begin{equation}
	\bf a^\star\rm = \delta L^\star/(\delta\tau^\star)^2=(\alpha/\gamma)\delta L/(\alpha/\gamma)^2(\delta\tau)^2=(\gamma/\alpha) \bf a\rm
\end{equation}

	The application of the mass-metric theory goes like this.  One writes the equations of motion within gravity classically, i.e. one uses Newton's law of universal gravitation.  He then applies the metric adjustments appropriate to the presence of gravity.  For a moving mass, he writes $(\gamma/\alpha)m_0$, for length he writes $(\alpha/\gamma)r$, etc.  When appropriate, he chooses $R=\infty$ so that $\alpha/\gamma=\alpha^2$ or, for fixed objects, sets $\gamma=1$.  If the problem involves orbital motion, one must use (10) and deal with the complexities this introduces.  The solution he obtains includes the effects of gravity, but is couched in terms of the gravity-free metric.
                                                                                        
\section{The Shapiro time-delay
}

	A critical test for any theory of gravity is the time delay of light, when the  light passes through a space influenced by gravity.  GR solves this problem by summing, over the path, proper time less clock time.  The mass-metric theory subtracts the time of passage at c from the time of passage at $c^\star$.  Shapiro and others (ref 4,5) have discussed the GR basis for this time delay and its experimental measurement.  For a radar signal from earth passing the sun on its way to a transponder on Mars, and for a return trip of the transponder radar signal, the delay as measured is up to about 248 microseconds when the beam grazes the sun's surface.  This conforms to the GR prediction, within 0.5 \%.

	Using the mass-metric formulation, one calculates the time of flight over the earth-Mars distance and back, using $c^\star$ as in (19) and also using c.  The difference is the delay due to gravity.
\begin{equation}
	 \delta t=-L/c+\int_L(1/c^\star)dx                                                                                                                                                   
\end{equation}
By L is meant the total round trip distance, 2L1+2L2, where L1 is the earth-sun distance (nominally $1.496\times10^{11}\hspace{0.04 in} m)$ and L2 is the sun-Mars distance (nominally $2.28\times10^{11}\hspace{0.04 in} m)$.  The radius of the sun is "a". 
\begin{equation}
	 \delta t=-L/c+(1/c)\int_L\alpha^{-2}dx=-L/c+(1/c)\int_L(1+GM/rc^2)^2dx                                                                                                                                                  
\end{equation}	                                                                                                                          
\begin{equation}
	\delta t\approx(2GM/c^3)\int_L dx/(x^2+a^2)^{.5}                                                                                                                                                  
\end{equation}
The result may be had by numerical integration of (23). \begin{equation}
	\delta t=(4GM/c^3)[\int_0^{L1}dx/(x^2+a^2)^{.5} +\int_0^{L2}dx/(x^2+a^2)^{.5}]                                                                                                                                              
\end{equation}
 Or, the integration can be had analytically, yielding t1.  	                               
\begin{equation}
	 t1=(4GM/c^3)[\ln((L1+(L1^2+a^2)^{.5})/a)+ \ln((L2+(L2^2+a^2)^{.5})/a)]                                                                                                                                            
\end{equation}	                                                                                 
This was done both ways, as there was some confusion about the exact form of the integral in (23).  The  integral (25) is from Vade Mecum.  (ref 6)

In either case, the resulting delay is 247.3 microseconds, and agrees with the measured time delay of 248 microseconds.  This experimental confirmation is critically important, and confirms that space itself seems to be expanded in the presence of a gravitational potential and, of course, exhibits non-Euclidean geometry.

	The analytical form of the time delay given by Shapiro looks different from (25).  The numerical value is the same in each case.  Shapiro writes the GR time delay as
\begin{equation}
	  \delta t=(4GM/c^3)ln((L1+L2+L)/(L1+L2-L))                                                                                                                                               
\end{equation}
The GR and mass-metric results however agree analytically, to first order, where the time delay, in each case, becomes:
\begin{equation}
	 \delta t=(4GM/c^3)ln(4L1L2/a^2)                                                                                                                                             
\end{equation}
 That these two different theories of gravity should yield the same numerical value and even the same analytical form for the correction to Newtonian gravity is interesting. Neither theory contains adjustable parameters.  

\section{The rate of precession of the perihelion of Mercury
}

	The rate of advance, of the perihelion of Mercury, is mostly a consequence of the motion of other planets.  Corrected for these, a residual rate of advance of 43.11 seconds per century remains and requires an explanation.   GR predicts 43.03 seconds per century, and this agreement is widely regarded as one of the strongest "proofs" of its validity.  (ref 7)

Newtonian gravity, with mass constant, predicts that the orbit of a planet about the sun will be an ellipse, an ellipse that does not precess.  The equation for an ellipse is
\begin{equation}
	 d^2u/d\theta^2 + u = A                                                                                      
\end{equation}                                                                           where u=1/r and A is
\begin{equation}
	 A = m/[\mu a(1-e^2)]\approx 1/[a(1-e^2)]                                                                                           
\end{equation}                                                                           The reduced mass $\mu$ for Mercury in the presence of the sun is essentially the mass of Mercury itself, m. It may be useful to define the parameters of the orbital path.  The semi-major axis length is "a" and the eccentricity is "e".  This means that the distance from the sun, at aphelion, is a(1+e) and at perihelion it is a(1-e).   
\begin{equation}
G=6.672 x 10^{-11}\hspace{.04 in} \end{equation}
\begin{equation}                                                                         
e=0.2056
\end{equation}
\begin{equation}                                                                                                                                                                   
M (sun) = 1.99 x 10^{30} \hspace{.04 in}kg
\end{equation}
\begin{equation}                                                                         
a=0.3871\hspace{.04 in} A.U.
\end{equation}
\begin{equation}                                                                                                                                                      
1 \hspace{.04 in}A.U. = 1.495 x 10^{11} \hspace{.04 in}meters
\end{equation}
\begin{equation}                                                                                                                                                    c=2.997925 x 10^8  \hspace{.04 in}meters/sec
\end{equation}
\begin{equation}                                                                                                                               
T=0.2408\hspace{.04 in} years \hspace{.04 in}is\hspace{.04 in} Mercury's \hspace{.04 in}period                                               
\end{equation}

The solution to the linear differential equation (28) is
\begin{equation}
	u = A(1+C\cos\theta)                                                                               
\end{equation}
as is easily shown by substitution.  "C" is the same as "e", and the resulting solution describes an orbit having natural angular frequency of "one".  This returns u to its perihelion value every $2\pi$ radians of rotation of the planet about the sun.  There is no precession in this case.  

	The calculation of the rate of precession by GR is complex.  GR modifies the differential equation (28), by adding a "$u^2$" perturbation  to A:
\begin{equation}
	A \rightarrow A+u^2(3GM/c^2)                                                                          
\end{equation}
The resulting differential equation is non-linear and its solution is difficult.  (ref 7)  After many pages of calculations and approximations, one eventually finds that $\theta$ must pass through $(1+\phi)2\pi$ radians between returns to the perihelion.  This advance of the perihelion, for a single orbit, is 
\begin{equation}
	2\pi\phi=2\pi(3GMA)/c^2                                          
\end{equation}
Expressed in seconds of arc per century, the rate is 43.03.  This agreement, with no adjustable parameters, is impressive.

	The mass-metric theory of gravity provides a much simpler way to calculate this rate of orbital precession.  It also allows one an intuitive understanding of \it why \rm there should be a precession.  In essence, the reason is that the changing mass, $\gamma m_0/\alpha$, at different distances from the sun, modulates the Newtonian gravitational force.  Since this increase of force is greatest at closest approach, we must expect an advance of the perihelion with each orbit.  
We first examine the metric of "u" and of "$d^2u/d\theta^2 $" and of "A", relative to the metric of an observer in zero gravity.
\begin{equation}
	u^\star \rightarrow (\gamma/\alpha)u
\end{equation}                                                                             
\begin{equation}
	d^2u^\star/d\theta^2 \rightarrow(\gamma/\alpha)(d^2u/d\theta^2)                                                              
\end{equation}
The constant A can also be written as $A=-(m/L^2u^2)(-GMm/r^2)$.  Simplifying,
\begin{equation}
	A=GM/r^2v^2
\end{equation}                                                                                      
\begin{equation}
	A^\star=GM/[(\alpha/\gamma)r(\alpha/\gamma)v]^2                                                                           
\end{equation}
\begin{equation}
	A^\star=(\gamma/\alpha)^4A
\end{equation}
Substituting these changes of metric, 
\begin{equation}                                                                                                                                            
	(\gamma/\alpha)d^2u/d\theta^2 + (\gamma/\alpha)u = (\gamma/\alpha)^4A
\end{equation}                                                                 
\begin{equation}
	d^2u/d\theta^2 + u = (\gamma/\alpha)^3A                                                                       
\end{equation}
We next expand $\gamma/\alpha$, neglecting terms higher than first order.
\begin{equation}
	 \gamma/\alpha=(1+(GM/c^2)(1/r-1/R))(1+GM/c^2r)\approx 1+(GM/c^2)(2/r-1/R)
\end{equation}
\begin{equation}
	(\gamma/\alpha)^3\approx1+6GM/c^2r-3GM/c^2R                                                                  
\end{equation}
\begin{equation}
	\delta\equiv 6GM/c^2                                                                  
\end{equation}
And so, 
\begin{equation}
	(\gamma/\alpha)^3A \rightarrow A((1+u\delta)-3GM/c^2R)
\end{equation}
\begin{equation}
	d^2u/d\theta^2 + (1-A\delta)u = A(1-3GM/c^2R)                                                                     
\end{equation}
Except for a small change in the constant term, the perturbation is "u" times $6GMA/c^2 (=A\delta)$ and can be compared with the GR perturbation, $u^2$ times ($3GM/c^2$).

The new differential equation (51) is linear, and is much easier to solve than the nonlinear GR differential equation (28) modified by (38).  Since we expect a precession, we consider a trial solution in the form
\begin{equation}
	u = A[1-3GM/c^2R](1+C\cos(1-\phi)\theta)                                                                       
\end{equation}
With this, u returns to its perihelion value when $(1-\phi)\theta=2\pi$.        
\begin{equation}
	du/d\theta =-A[1-3GM/c^2R]C(1-\phi)\sin(1-\phi)\theta                                                             
\end{equation}
\begin{equation}
	d^2u/d\theta^2 =-A[1-3GM/c^2R]C(1-\phi)^2 \cos(1-\phi)\theta                                          
\end{equation}
Substituting and collecting terms,
\begin{equation}
	(A[1-3GM/c^2R]C\cos(1-\phi)\theta)(2\phi-\phi^2-A\delta)=A^2[1-3GM/c^2R]\delta
\end{equation}
To first order, $\phi^2$ vanishes, and the advance of the perihelion each orbit is
\begin{equation}
	2\pi\phi= (A\delta/2)2\pi =(3GMA/c^2)2\pi                                                          
\end{equation}
in radians, a finding identical with the GR result, (39).  As with GR, there are no adjustable parameters.

	Of course, this trial solution requires that $A^2\delta$ should vanish, which it does not.  We can modify the trial solution (52) to take care of this,
\begin{equation}
	u = A[1-3GM/c^2R](1+A\delta)(1+C\cos(1-\phi)\theta)                                                           
\end{equation}
Substituting in the differential equation, dividing through by $[1-3GM/c^2R]$ and collecting terms,
\begin{equation}
	A(1+A\delta)(C\cos(1-\phi)\theta)(-(1-\phi)^2+1-A\delta]=A-A(1-A\delta)(1+A\delta)
\end{equation}
Keeping only first order terms in "$\delta$", the orbital offset of the perihelion remains the same but the constant term now vanishes.  It should be noted that the actual value of "R" for this bound orbit is of little significance.  In an earlier version of this paper, R was simply taken as infinite and the calculated rate of advance of the perihelion is identical with (56).  The mass-metric formulation simplifies the calculation enormously, by requiring only the solution of a linear differential equation, (51).  Only one who has gone through the agony of solving the nonlinear GR differential equation (28) modified by (38) can fully appreciate the difference.

\section{The deflection of starlight
}

	As for the deflection of starlight, the inferred reduction of speed (19) relative to an outside observer means that gravitationally influenced space has an index of refraction: 
\begin{equation}
	n=c/c^\star=1/\alpha^2\approx 1+2GM/c^2r                                                                
\end{equation}
The deflection of starlight is entirely accounted for by refraction, using Snell's law.  Using a small angle approximation throughout, consider two light paths grazing the sun, parallel to the x axis.  One is along the x axis and the other is at a small height "h" above the x axis.  The origin at x=0 is a distance "a" above the center of the sun.  The light path is $-\infty$ to $\infty$, along "x".  Over a distance dx, the difference of optical path lengths is 
\begin{equation}
	\delta x=dx[1+GM/c^2r]^2-dx[1+GM/c^2(r+h\cos\theta)]^2\end{equation}
\begin{equation}
	\delta x\approx dx(2GM/c^2r^2)h\cos\theta \mbox{  where  }
  \cos\theta=a/r \hspace{.1 in} and \hspace{.1 in} r=\sqrt (x^2+a^2)                                                  
\end{equation}
The wavefront of the light is tilted by $d\phi$:
\begin{equation}
	d\phi\approx \delta x/h=dx(2GM/c^2r^2)cos\theta= dx(2GM/c^2r^2)a/r
\end{equation}	
On performing the integration over all "x", the deflection is
\begin{equation}
	\phi=4GM/c^2a                                                                                    
\end{equation}
and agrees with the GR calculation.  (ref 8)

\section{Re-examining the effect of gravity on rest mass}

	As mentioned at the end of section 2, the assumption that the gravitational red shift of a photon occurs gradually seems to be in error.  There can be no gravitational force on a photon, since we can fully account for the deflection of starlight by refraction. So, gravity has not the power to change the energy of a photon.  The observed gravitational red shift hence occurs abruptly, at the moment of formation of the photon.  

	We regard (9) as validated by its successes in dealing with the classical tests of GR.  Consider again the gedanken annihilation experiment.  A rest mass $m_0$ at the upper level annihilates, and the EM radiation energy is $m_0c^2$.  Assume that $\alpha=1$ at the upper level.  At the lower level, the rest mass (9) is $m_0/\alpha$ but the energy is only $\alpha m_0c^2$.  We are forced to the conclusion that annihilation of this rest mass yields only this energy, $\alpha m_0c^2$.  So, gravity increases rest mass but lowers the total energy (including potential energy).  The literature on this point is ambiguous.  French (ref 9) says that a charged capacitor has more mass than an uncharged capacitor.  It certainly has a greater energy content.  By the arguments given here, there should be no increase of mass upon charging a capacitor.  That is, mass is a consequence of kinetic energy and changes of potential energy do not affect mass.  Annihilation energy measures total energy contained in the particle, and is not always $m_0c^2$.  Experimental verification of this point is badly needed.

\section{The gravitational red shift}

	The gravitational red shift is concerned with photons, photons which are created in a gravity environment and which emerge with less energy than might otherwise be expected.  The mass-metric contraction of time suggests that these photons should have a higher frequency, and the finding that gravity exerts no force on a photon means that these photons should emerge with this higher frequency.  This seems, off-hand, enigmatic.

	The solution lies with a consideration of the energetics.  Consider an excited atom, perhaps a hydrogen atom in a 2s state.  When this de-excites, in the absence of gravity, a Lyman alpha photon is loosed and $E=h\nu_0$.

	Case I:  The excited atom is released far from a massive body, M, and it moves around M in a hyperbolic orbit.  The total energy remains constant, with the gain of kinetic energy balanced by the loss of potential energy.  Somewhere during the orbit, the Lyman alpha photon is released.  When detected, away from the gravitational environment, its energy is still $E=h\nu_0$.  By the mass-metric view, the period of the photon at release is shorter,
 $\tau=(\alpha/\gamma)\tau_0=\alpha^2\tau_0$.  
Yet, the energy is unchanged. Since the frequency is larger, $\nu=\nu_0/\alpha^2$, $E=h\nu_0=\alpha^2h\nu$.
Energetically, there is a gravitational red shift of $\alpha^2$.

	Case II:  The excited atom is stopped somewhere during its orbit of M, and the kinetic energy is converted to heat and lost.  $\gamma$ becomes equal to one.  The released Lyman alpha photon  will have less energy, when detected away from the gravity environment, $\tilde{E}=\alpha E=\alpha h\nu_0$.
By the mass-metric concept, $\nu=\nu_0/\alpha$.  Again, we find that $\tilde{E}=\alpha^2 h\nu$.

	And so, energy considerations impose a red shift on photons arising in a gravity environment and detected without.  This red shift is $\alpha^2$.  In other words, the energy of a photon produced in a gravity environment is less than $h\nu$, by the factor $\alpha^2$.  In a gravity environment, $E=\alpha^2 h\nu$ instead of the usual $E=h\nu$.
 
\section{A hole of a different color
}

 This failure to find a force of gravity on photons  conflicts with one of the findings of GR, the event horizon, generally interpreted to mean that the tug of gravity exerted by a black hole is so great as to prevent even the escape of light.  There is a perverse symmetry in the disagreement between the GR and mass-metric viewpoints.  Although GR considers motion in a gravitational field be force-free along a geodesic, the concept of force on photons is suggested to explain the behavior of a black hole.  The mass-metric formulation calls for real forces to explain the motion of a mass in a gravitational potential, yet denies that gravity exerts a force on photons.

Although the speed of light for an outside observer seems diminished by $\alpha^2$, it can never become zero or imaginary as one might expect for "trapped light".  It is clear that light is deflected by changes of the gravitational potential, and so one might inquire, as an alternate trapping mechanism, into whether or not a circular path of light is possible at some radius.  Perhaps the easiest way to look at this is to consider the optical length, nL, where "n" is the index of refraction (59) and L is the path length, $2\pi r$.  For a circular path, nL must be an extremum, and so the derivative of nL with respect to r must vanish.  Since this derivative is positive definite, no such path is possible. 
 
We know that an approaching photon's path is bent towards the mass responsible for the gravitational potential.  Depending on its impact parameter, "a", i.e. the distance it would miss the mass if it were to travel along a straight line, the position of closest approach will be $a^\star=\alpha^2a$.  At closest approach, the path is perpendicular to a radius vector.  The path then proceeds outward, unless $a^\star$ is less than the radius of the mass in which case the photon is absorbed or reflected.  The principle of optical reversibility suggests that any of these light paths is reversible.  If reflected, the photon takes an outward path and eventually is released by the gravity field.  If absorbed, one expects some degradation of the energy followed by re emission of several lower energy photons.  These also will follow an outbound course and eventually escape. 

This suggests that mass can be captured, permanently, but not light.  A sufficiently massive body can store up large quantities of light for later release.  Whatever of this light is absorbed by the mass, and later emitted thermally, will be red shifted.  Any light which misses the central mass, or which is reflected off the central mass, will eventually emerge with its original energy.  Such a body might be termed a red hole, but not a black hole.  Depending on just "how red" such a body is, it might be called "dark matter."

The possible existence of black holes is an old concept.  The logic seemed reasonable, that a mass m lies in a potential like (11) and its potential energy becomes negative without limit as "r" approaches zero.  Escape from this potential well is only possible if the kinetic energy of a particle exceeds the negative potential energy, and $1/2\hspace{0.03 in} mv^2$ has an upper limit of $1/2 \hspace{0.03 in}mc^2$.  If one sets this condition, which seemed reasonable before SR,
\begin{equation}
 	1/2\hspace{0.03 in} mc^2=GmM/r                                                                                                                                                                    
\end{equation}
it follows that escape is impossible for any "r" less than the Schwarzschild radius, $2GM/c^2$.  Even after SR, the maximum kinetic energy seemed to be $m_0c^2$, again suggesting the possibility of black holes.  In rebuttal, repeating (12) and (13), the mass-metric formulation finds for the potential energy
\begin{equation}
	U=(1-1/\alpha^2)m_0c^2 \approx-2Gm_0M/r                                                             
\end{equation}
and the kinetic energy 
\begin{equation}
	T=(1/\alpha^2-1)m_0c^2\approx 2Gm_0M/r                                                            
\end{equation}
for a body falling freely from a far distance.  This means that T+U=0 for every radius, and there is no reason to expect "black holes".

\section{The Fifth Force ?}

	A number of experiments have been conducted, in an attempt to find either a composition-dependent force of gravity or a deviation from inverse square behavior.  Will (10) summarizes these, and considers them disproved by the data.  The mass-metric approach of this paper suggests that inverse square is not wholly correct.

	For orbital mechanics, the gravitational force (for R large) is
\begin{equation}
	F=-(\gamma/\alpha)^3Gm_0M/r^2\approx -(Gm_0M/r^2)(1+6GM/rc^2)                                                          
\end{equation}
If one measures the static force, $\gamma=1$ and the gravitational force is 
\begin{equation}
	F=-(1/\alpha)^3Gm_0M/r^2\approx -(Gm_0M/r^2)(1+3GM/rc^2)                                                          
\end{equation}
Since Will concludes that the null results of "fifth force" experiments place an upper limit on a fifth force of $10^{-3}$ to $10^{-6}$ times "g", one can ask how much deviation from "inverse square" may one expect from this mass-metric approach.  The ratio of the force in (68) to the classical Newtonian force is $1+3GM/rc^2$.  Evaluating this ratio at the surface of the earth, it differs from one by $2*10^{-9}$ which is three orders of magnitude smaller than the limit of the null experimental results. The nominal sensitivity of gravimeters is at the $10^{-9}$ g level, and continues to improve. This modification of Newtonian gravity (68) may soon be thought important for the interpretation of gravity surveys.

	The precession of the perihelion of Mercury can be understood as a direct consequence of this deviation from an inverse square law.  Note that the deviation is in that case twice as large (67) as one expects for the static case (68).  Although this "fifth force" seems to be real, in the sense that gravity is not inverse square, it is in no sense a new fundamental law of nature.

\section{Curvature of space}

	The concepts of curvature and non-Euclidean geometry of space, in the presence of a massive body, arose from Einstein's GR theory.  The mass-metric theory tends to support these concepts, but the explanation differs.  In the present theory, empty space near a massive body is intrinsically unknown.  It is only when one measures its properties that one can conclude whether the geometry is Euclidean, or not.  When measured, we find that meter sticks shrink, mass increases, clocks run differently, and light travels along curved paths.  So, the measured space is non-Euclidean.  Although we cannot physically measure empty space, we can infer something about it.  We have postulated that empty space is Euclidean, and calculated the consequences using the mass-metric theory.  If these consequences were  that the measured geometry is Euclidean, then we could be sure that empty space is non-Euclidean.  On the other hand, since the measured consequences agree precisely with the calculated consequences, we are assured that our postulate was correct and that empty space is indeed Euclidean.  It appears that empty space really is Euclidean, whether there be gravity or not.

	All this is rather disheartening for an experimental physicist.  It finds that his measuring instruments change calibration in a changing gravitational potential.  Ironically, this means that an experimental  finding of non-Euclidean geometry in the presence of a massive body can be used to confirm that empty space is indeed Euclidean.

	An imperfect analogy may be useful for understanding the foregoing arguments.  Suppose that we conduct experiments in a room in which a temperature gradient is maintained.  The hot spot is compact, and the temperature falls off with distance from that hot spot.  An ordinary meter stick changes length, according to temperature.  $L=L_0(1+bT)$  If we were measuring the Coulomb force of one charge on another, we would find that it was not quite inverse square.  The reason is obviously that the meter stick changes calibration, at different distances from the hot spot.  However, if we recognize the effect of temperature on the length and make corrections to account for the effect of temperature, we then find that the Coulomb law is inverse square.  If we were unaware of the temperature effect on the metric as measured by the meter stick, we might easily be led to believe that the geometry inside the room was not Euclidean.  A circle might be drawn about the hot spot, and the diameter and the circumference measured using meter sticks.  The result must be that $C < \pi D$. We might decide that Riemannian geometry was needed to properly describe events within this room, requiring the use of a metric tensor.  Another observer, located outside the room, might understand the effects of  temperature and how it affects the length of a meter stick.  He would then adjust the measured distance for the effect of temperature, and his conclusions (from the same data) would include that the Coulomb law is indeed inverse square and that $C = \pi D$.  He would have no reason to doubt that the space within the room was Euclidean.

	In conclusion, then, it is here suggested that gravity increases the rest mass of an object.  Taken together with the SR boost of mass due to motion, the "outside" observer concludes that mass increases with speed and with gravity.  This increase of mass shortens meter sticks and speeds up clocks, in well-defined ways.  On accounting for this change of metric, he finds that Newtonian gravity works very well indeed.  He sees no need to invoke non-Euclidean geometry.  He finds no need to invoke curved space-time to account for the consequences of experiments.

\section{Discussion
}

One of the key elements in finding this mass-metric formulation of gravity was the Pound-Rebka experiment, which can be interpreted to show that rest mass increases under gravity.  It grates on one's sensibilities to recognize that the  loss of conventional potential energy, $U= -GmM/r$ for a falling body, is only half the gain of kinetic energy.  The gain of kinetic energy is in two parts; one is the usual $1/2\hspace{.03 in} mv^2$ and the other is $\delta m_0c^2$.  The latter kinetic energy lies wholly within an elementary charged particle, and manifests itself as an increase of rest mass.  The $1/2\hspace{.03 in} mv^2$ kinetic energy is exchangeable and can be degraded to heat, but the $\delta mc^2$ internal kinetic energy cannot.  Although the Pound-Rebka experiment is now 40 years old, its implications for a gravitational increase of rest mass seem to have escaped notice. 

	The difference of clock rates in the GR and  mass-metric theories is more apparent than real.  The effect of gravity in the mass-metric theory is to speed up the clock rate, which shortens the inferred time between ticks of a \it stationary \rm clock by $\alpha$.  The gravitational red shift, for an observer outside this gravitational potential, reduces the energy of photons by $\alpha^2$. Since the rate of an atomic clock responds to the energy of photons, this slows down slows the perceived rate of the clock by a factor $\alpha^2$.  The net result is that the outside observer observes the gravitationally-influenced clock running slow, by the factor $\alpha$, in agreement with GR. (ref 11)

 	The Shapiro time delay experiment is extremely important.  It confirms the apparent reduction of the speed of light, and supports the explanation of the gravitational bending of light as simply a matter of refraction.  Taken together, they show that gravity exerts no force at all on photons.  

	The complications inherent in self-gravitational effects may require attention in some cases.  The rest mass of a hydrogen atom anywhere on or within the earth exceeds that of a hydrogen atom in outer space.  So, to be consistent, we should correct the mass of, say, the earth by giving each element a $1/\alpha$ boost because of the self-gravitational potential.  

	Some have speculated that G might be slowly changing over time.  If one insists on using the original Newtonian formulation to define G, in an expanding universe, the mass-metric formulation shows that the average gravitational potential is decreasing as masses move apart.  This effect could be absorbed within G, requiring that G be slowly decreasing.  I prefer to hold G constant, and recognize that all masses in the universe are slowly decreasing, on average, because the average gravitational potential is declining.  That is, aside from any net mass loss by radiation of EM energy, the total mass of the universe is decreasing in consequence of the Hubble expansion.

	It may be useful to reexamine the basic treatment of gravity by GR in light of the variations of mass with velocity and especially with gravitational potential.  Duffey (ref 12), in discussing gravity in GR, specifically denies that rest mass changes with gravitational potential and describes potential energy like (11) instead of (13).  The "event horizon" in GR may turn out to disappear when mass is treated like (10) and when the gravitational potential is correspondingly modified.

\section{Conclusions
}

	Newtonian gravity, in its original form where mass was thought  constant and the geometry of space Euclidean, accounts for gravitational phenomena in the limit of weak gravity.  Very accurate experimental results have found small discrepancies, fully explicable (without adjustable parameters) until now only by GR, that indicate Newton's law is not quite right.  Scientists have had to revise, over time, some of their most cherished "conservation" concepts, such as energy, to include other forms of energy such as heat and mass.  The flip side of conservation of energy is that mass alone is not conserved.  Once one recognizes that mass changes according to motion and the gravitational potential in which it sits, and uses this variable mass in Newton's law of universal gravitation, Newtonian gravity conforms with experiment on an equal footing with GR.  And, it is much easier to calculate the effects!  

	Aside from the ongoing controversy about black holes, the consequences of these two theories appear to agree.  As for the ease of computation using the mass-metric scalar formulation of Newtonian gravity, physicists have long since discovered that scalar fields are easier to work with than vector fields.  

	Finally, Einstein's concept of gravity seems to have been "gravity causes space to curve, and the curved space tells mass how to move."  The present paper presents a different interpretation.  Gravity causes mass to increase, and the increase of mass makes measured space seem curved.  The nature of empty space near a massive body is intrinsically unknown until its properties are measured, and the mass-metric theory indicates that measurements do not correctly reflect the nature of space.  As measured, space seems non-Euclidean and this appears to support "curved spacetime". However, since these changes are well defined, one can apply an inverse transformation to the non-Euclidean measured space and can conclude that empty space is everywhere Euclidean.

	GR has so insinuated itself into the way physicists think about gravity that alternate theories of gravity are now judged in terms of GR concepts.  Will (ref 13) says, "The Einstein Equivalence Principle is the foundation of all curved spacetime or 'metric' theories of gravity, including general relativity."  Further, "This (EEP) principle is at the heart of gravitation theory, for it is possible to argue convincingly that if EEP is valid, then gravitation must be a curved-spacetime phenomenon, i.e., must satisfy the postulates of Metric Theories of Gravity.  These postulates state: (i) spacetime is endowed with a metric \bf g \rm,..."
The mass-metric theory of gravity lies  outside Will's carefully crafted language for the comparison of theories of gravity.  The fact that measured spacetime is curved is here shown to be an artifact of the measurement, attributable to the changing calibration of instruments according to the gravity in which they are sited.

\section{References
}

\hspace{0.22 in}1.  Pound, R.V. and Rebka, G.A. Jr., Phys. Rev. Letters \bf 4\rm, 337-41 (1960).

2.  Collins, R.L., An electromagnetic basis for gravity, url:\\http://publish.aps.org/eprint/gateway/eppackage/aps1999jun08\_001.

3.  Halliday, D. and Resnick, R., \it PHYSICS\rm \hspace{.05 in}(John Wiley \& Sons, NY 1978), 1107-8.

4.  Shapiro, I.I., Phys. Rev. \bf 141\rm, 1219-22 (1966).

5. Shapiro, I.I., et. al., J. Geophys. Res. \bf 82\rm, 4329-34 (1977).

6. Anderson, H.L., ed. in chief, \it Physics Vade Mecum \rm(AIP, 1981), 15.

7. Scott, T. and Madore, B., Perihelion precession rate of Mercury according to General Relativity, url:  http://www.maplesoft.com/cybermath/html/mercury.html

8. G.H. Duffey, \it THEORETICAL PHYSICS\rm \hspace{.05 in}(Houghton Miflin Co., Boston, 1973), 597.

9.  A.P. French, \it SPECIAL RELATIVITY \rm, \hspace{.05in}(W.W. Norton and Company, New York, 1968), 19.

10.  C.M. Will, \it THEORY AND EXPERIMENT IN GRAVITATIONAL PHYSICS \rm \hspace{.05in}(University Press, Cambridge, 1993), 341-3.

11. ibid, 599.

12. ibid, 590.

13. ibid, 22.
  
% Enter section title between curly braces

\end{document}